\documentstyle[eqsecnum,preprint,aps,epsfig]{revtex}

\tightenlines

\newcommand{\cs}{\'{c}}

\newcommand{\beq}{\begin{equation}}
\newcommand{\eeq}{\end{equation}}
\newcommand{\bdm}{\begin{displaymath}}
\newcommand{\edm}{\end{displaymath}}
\newcommand{\beqa}{\begin{eqnarray}}
\newcommand{\eeqa}{\end{eqnarray}}
\newcommand{\beqab}{\begin{eqnarray*}}
\newcommand{\eeqab}{\end{eqnarray*}}
\def\nn{\nonumber}

 \def\@makefnmark{\hbox to 0pt{$^{\@thefnmark}$\hss}}  

\def\M4{\overline M}
\def\R4{\overline R}

\newcounter{saveeqn}%

\begin{document}

\draft


\preprint{NYU-TH-00/02/01, HEP-MN-1940}

\title{The Power of Brane-Induced Gravity}

\author{Gia Dvali$^1$ \footnote{e-mail:gd23@nyu.edu}, Gregory Gabadadze$^2$
 \footnote{e-mail: gabadadz@physics.umn.edu}
Marko Kolanovi\cs$^1$ \footnote{e-mail: mk679@nyu.edu}, Francesco Nitti$^1$ 
\footnote{e-mail: fn230@nyu.edu} }
\address{$^1$Department of Physics, New York University,
New York, NY 10003\\
$^2$Theoretical Physics Institute, University of Minnesota, 
Minneapolis, MN, 
55455}
\date{\today}
\maketitle
\begin{abstract}
We study the role of the brane-induced
graviton kinetic term in theories with  large extra dimensions.
In five dimensions we construct a model with a TeV-scale fundamental
Planck mass and a {\it flat} extra dimension
the size of which can be astronomically large.  4D gravity on the brane
is mediated by a massless zero-mode, whereas the couplings of the
heavy Kaluza-Klein  modes to ordinary matter are suppressed.
The model  can manifest itself through the predicted
deviations from Einstein theory  in  long distance precision
measurements of the planetary orbits. 
The bulk states can be a rather
exotic form of dark matter, which at sub-solar distances interact
via strong  5D gravitational force. 
We show that the induced term 
changes dramatically the phenomenology of 
sub-millimeter extra dimensions. For instance,
high-energy constraints from star cooling or cosmology 
can be substantially relaxed.

\end{abstract}

\narrowtext
\newpage

\section{Introduction}
\setcounter{equation}{0}

It has been known for some time \cite {ADD,ADD2} that in theories with large 
extra dimensions the scale at which gravity becomes strong, 
$M$, can be much lower than the four-dimensional Planck mass $M_P$.
In this approach the standard hierarchy problem takes a different meaning 
since the field theory cut-off is lowered. The four-dimensional 
gravity is weak due to the fact that gravitational flux spreads into $N$ 
extra dimensions. The relation between fundamental and observed Planck 
scales takes the form:
\beq
M_P^2 ~= ~M^{2 + N}~ R^N~,
\eeq
where $R$ is the circumference of the extra dimensions. 
In this framework the Standard Model  
particles are localized on a brane whereas gravity propagates in the bulk 
of $N$ new dimensions. Until now, it was assumed that if $M\sim 
{\rm few~ TeV}$ the $N=1$ case is ruled out by observations since 
it would require  $R  \sim 10^{16}$ cm (the solar system size).
As to the higher codimensions, they allow $R$ 
to be around a millimeter or smaller. 
The purpose of the present paper is to argue 
that the framework of compact flat extra dimensions can be dramatically modified due to 
the presence of the  brane-induced kinetic term for higher dimensional
graviton \cite{DGPind,DG}
 
\beq
S_{\rm ind}~\propto~\int d^{4}x~
\sqrt{|{\rm det}{\bar g}|}~{\R4}(x)~,
\eeq
where $\bar g$ is the higher dimensional metric evaluated at the position
 of the brane, and ${\R4}(x)$ is the corresponding four dimensional Ricci scalar. As a result of this term the flat extra dimension can be astronomically large, and many high energy constraints can be lifted. 

In fact, in the approach of \cite{DGPind,DG}, the extra 
dimension was taken to be truly infinite.
In spite of this, it was shown 
that the model reproduces  four-dimensional gravity due to 
the four-dimensional curvature term in the brane worldvolume theory.

To be more precise let us consider $5$-dimensional Minkowski space with a 
standard bulk gravitational action
\beq
S_{\rm bulk }~=~\int d^{4+1}X~\sqrt{|G|}~{\cal L}\left
(G_{AB}, ~{\cal R }_{ABCD}, ~\Phi\right )~,
\label{act1}
\eeq
where the capital Latin indices run over $D=(4+1)$-dimensional space-time.
$G_{AB}$ denotes the metric of 5-dimensional
space-time, ${\cal R}_{ABCD}$ is the 5-dimensional
Riemann tensor and $\Phi$ collectively denotes other fields. 
Suppose that there is a
3-brane in this space. The  3-brane can be realized as a soliton
of the corresponding field equations. 
We split the coordinates in 5-dimensions as follows:
\beq
X^A~=~(x^{\mu},~y)~,
\eeq
where Greek indices run over four-dimensional brane worldvolume,
$\mu=0,1,2,3$~, and $y$ is the coordinate transverse to the brane.
In order to reduce our discussion to its main point the brane will 
be taken to have  zero width\footnote{This is a good approximation
when the mass of the field (presumably a scalar) out of which 
the brane is made is bigger than the energy scale of the low-energy 
effective theory.}. In this approximation the brane action takes 
the form: 
\beq
S_{\rm 3-brane}~=~-T 
~\int d^{4}x~\sqrt{|{\rm det} {\bar g}|}~,
\label{brane0}
\eeq 
where $T$ stands for the brane tension and 
${\bar g_{\mu\nu}}=\partial_\mu X^A\partial_\nu X^B G_{AB}$ 
denotes  the induced metric on a brane\footnote{Below 
we will imply that the Gibbons-Hawking surface term is also
present on the worldvolume so that it provides the correct
bulk Einstein equations.}.
In  most  of this  work, unless otherwise stated,
we neglect brane fluctuations. Therefore, the  induced metric 
can be written  as follows:
\beq
{\bar g}_{\mu\nu}~(x)~=~G_{\mu\nu}\left (x, y=0 \right )~.
\label{gind}
\eeq
In general, there could  be localized matter fields on the brane 
worldvolume. These can be taken into account by writing the 
following action for the brane:
\beq
{\tilde S}^{\rm matter}_{\rm 3-brane}~=~S_{\rm 3-brane}~+~
\int d^{4}x~\sqrt{|{\rm det} {\bar g}|}~{\tilde {\cal L}} (\phi)~,
\label{brane1}
\eeq  
where $\phi$ collectively denotes all the localized 
fields for which the four-dimensional 
Lagrangian density is ${\tilde {\cal L}}$. 
In  the 
classical theory, which we are discussing so far, the
4D Ricci scalar on the brane worldvolume is not present. 
Thus, the localized particles separated at a distance $r$
on a brane interact via  the $(4+1)$-dimensional
gravitational force-law, that is $F\sim 1/r^{2+1}$.
This holds as long as the classical theory is concerned. 
However, in the full quantum theory the 4D Ricci scalar
will be generated (along with other terms) on the  brane worldvolume. 
This is due to quantum loops  
of the matter fields which are {\it localized} on the  brane \cite{DGPind,DG}. 
As a result, the following   worldvolume terms should be 
included when one considers   the full quantum theory:
\beq
S_{\rm ind}~=~{\M4}^2~\int d^{4}x~
\sqrt{|{\rm det}{\bar g}|}~\left [ {\bar \Lambda}~+~ {\R4}(x)~+{\cal O}
\left ( {\R4}^2  \right ) \right ]~,
\label{4DR}
\eeq  
where ${\M4}^2\equiv r_c~M^3$ is some parameter which depends 
on the details of the worldvolume model \cite {DGPind,DG}. 
$\bar \Lambda$  in (\ref {4DR}) is an 
induced four-dimensional cosmological 
constant. The role of this term is to renormalize 
the brane tension.  Furthermore, ${\R4}(x)$
is the four-dimensional Ricci scalar which is constructed 
out of the induced metric ${\bar g}_{\mu\nu}(x)$ 
defined in (\ref {gind}). In five-dimensional Minkowski space
a brane with nonzero tension inflates \cite{Vilenkin,Sikivie}.
Therefore, to avoid the worldvolume inflation we fine tune the 
brane tension $T$ and the brane worldvolume cosmological constant 
$\bar \Lambda$ so that the net tension is vanishing
\beq
T^\prime\equiv ~T-{\bar \Lambda}~\M4^2~=~0~. 
\eeq
This is a usual fine tuning 
of the four-dimensional cosmological constant.

The graviton propagator resulting from such a system is quite peculiar.
Ignoring the tensor structure for a moment we obtain for the 
corresponding Green's function the following expression:
\beq
{\tilde G}_R(p,y)~=~{1\over 2M^{3}p~+~\M4^2p^2}~\exp\{-p|y|\}~.
\label{G5}
\eeq
Here $p^2$ is a four-dimensional Euclidean momentum and $p\equiv\sqrt{p^2}$. For 
sources which are localized on the  brane ($ y = 0$), 
this propagator reduces to a massless 
four-dimensional Green's function
\beq
{\tilde G}_R(p, y=0)~\propto~{1\over p^2}~,
\label{G5massless}
\eeq
provided that $p\gg 1/r_c$. Thus, at distances $r\ll r_c$ 
we observe  the correct
Newtonian behavior of the potential  
\beq
V( r )~ \propto ~{m_1~m_2 \over r~\M4^2}~.
\eeq
At large distances, $r \gg r_c$, however, the behavior of the Green's
function changes 
\beq
{\tilde G}_R(p, y=0)~\propto~{1\over p}~.
\label{G5massless1}
\eeq
This gives rise to 
 a  Newtonian potential 
which scales in accordance with the laws of a five-dimensional theory
\footnote{ Transition to five dimensional regime at long distances is
also a characteristic feature of the models with "quasi-localized"
gravity \cite{GRS1,Csaki1,DGP1,DGP2,PRZ,CECH,Parikh}. However, these
models usually include negative norm states\cite{DGP1,DGP2,PRZ}.}

\beq
V( r )~\propto~{1 \over r^2~M^3}~.
\eeq
This somewhat puzzling behavior can be understood in two equivalent ways
which we briefly discuss.
First let us adopt the  five-dimensional point of view.
In this language, although there is no localized massless particle, 
there exists a localized resonance state in the spectrum.
The lifetime of this resonance is  $\sim 
r_c$. The resonance decays into the continuum of modes. 
This can be manifestly seen using the  
K\"allen-Lehmann representation for the Green's function
\beq
{\tilde G}_R(p,y=0)~=~{1\over 2M^{3}p~+~M^3r_cp^2}~ = ~
\int_0^\infty {\rho (s)~ds \over s~+~p^2 }~,
\label{KL}
\eeq
where the spectral density as a function of the Mandelstam variable $s$
takes the form:
\beq
\rho (s) ~\propto ~{1\over \sqrt{s}}~{r_c\over 4~+s~r_c^2}~.
\eeq
As $r_c\rightarrow \infty $ the spectral density 
tends to the Dirac function $\rho(s) \rightarrow const.\delta(s)$
describing a stable massless graviton (this corresponds to the limit 
when the bulk kinetic term can be neglected).
Summarizing, at distances $r < r_c$ the resonance mimics the massless 
exchange, and therefore  mediates the $1/r^2$ force. 
At larger distances, however, it 
decays into the continuum, and as a result  the 
force-law becomes that of a five-dimensional theory, $\sim1/r^3$.

A different but equivalent way to 
understand  the above result is 
to adopt the point of view of the four-dimensional mode expansion.
The analysis of the linearized equation for the small fluctuations shows 
(see Appendix) that there is a continuum of 4D massive states  with 
 wave-function profiles that, at the location of the brane, are 
suppressed by the following factor 
\begin{equation}
|\phi_m(y=0)|^2~ = ~{1\over 4~+ ~m^2~r_c^2}~.
\end{equation}
The Newtonian force on the brane is mediated by the exchange of all 
these Kaluza-Klein (KK) modes.  These give rise to the  expression:
\begin{equation}
V( r )~ \propto ~{1 \over M^3}~ \int_{0}^{\infty} {dm \over 4+ m^2r_c^2}
{e^{-mr}\over r}~.
\end{equation}
At any distance
$r$ the dominant contribution comes from the modes lighter than $m=1/r$.
The modes with $m < 1/r_c$ have 
unsuppressed wave-functions on the 
brane. Therefore, for  $r > r_c$ the interaction picture is similar 
to that of a  five-dimensional theory. 
In contrast, when  $r<r_c$ the picture changes 
since the modes with $m > 1/r_c$ have 
suppressed couplings. Although the  
number of the modes which  participate in the exchange
at a given distance $r < r_c$ is the same as 
in the five-dimensional picture, 
their contributions  are suppressed. Thus, the number of the light modes 
effectively contributing to the exchange "freezes-out" and the resulting 
behavior of the potential is $1/r$.

As pointed out in \cite{DGPind}, this model exhibits the van 
Dam-Veltman-Zakharov (vDVZ) discontinuity \cite{Veltman,Zakharov} 
in the tensor structure of the graviton 
propagator in {\it the lowest tree-level approximation}
(as would any flat space ghost-free theory where 
$4D$ gravity is mimicked by the exchange of (a continuum of) 
massive spin-2 particles \cite{DGP1}).
Another disadvantage of this model is the low value of the 
crossover scale  $r_c\simeq 10^{16}~ {\rm cm}$. 
This would imply that Newtonian gravity is modified at this distance,
which is certainly unacceptable.

In the present paper we show that both aforementioned difficulties 
are circumvented  if the 
fifth dimension is  compactified on a circle of  macroscopic size $R$.
We will argue that $R$ can be of astronomical size.
In this framework, the 
four-dimensional Planck scale 
emerges as
\begin{equation}
M_P^2~ = ~M^3~(~R ~+ ~r_c~)~.
\label{relation}
\end{equation}
Gravity on the brane is dominated by a lowest lying massless 
KK zero mode, which has an unsuppressed wave-function. 
As a result there is no vDVZ discontinuity problem.
Contributions from the massive KK modes give rise to 
small corrections  to Newtonian 
gravity. These corrections  are mostly 
constrained by the data on the motion of the inner planets, Moon and 
satellite experiments. 
We study constraints coming both from astronomy (planetary motion) and 
from astrophysics (star cooling), and show that the model is consistent 
with all the observations.

\section{The Model} 
\setcounter{equation}{0}

We would like to consider a simplest possible 
model with a single brane  embedded in 
five-dimensional space with one compact dimension.
The action of our model takes the form:
\beq
S~=~ M^3~\int d^4x~\int_{0}^{R}dy ~\sqrt {G}~{\cal R}_{(5)}~+~
M^3~r_c~\int d^4x ~\sqrt {|{\bar g}|}~{\R4}~.
\label{1}
\eeq
As we mentioned before, the four-dimensional 
Planck mass is set by the sum of the two distance scales $r_c$ and $R$ 
through the relation (\ref{relation}). 
For the solution of the hierarchy problem along the 
lines of Ref. \cite{ADD} we can put  $M \lesssim$ TeV. 
This sets the largest 
of the two scales, $R$ and $r_c$,  to be 
larger then $\sim 10^{16}$cm or so. 
It is tempting to assume that the origins of these two scales are related. 
Some possible ideas along this direction will be discussed later. For 
the time being we shall take them as the parameters of the theory.

We are interested in the Newtonian gravitational 
interaction of the brane-localized sources. 
In this section we will ignore the 
tensor structure of the graviton propagator 
and concentrate on its scalar part.
A straightforward calculation (see Appendix) 
leads to the following result for the Euclidean
two-point Green's function:
\begin{eqnarray} \label{propagator}
\Delta (p, 0)
&=& \frac{1}{p^2}\left[\frac{1}{1+ (1/pr_c)2\tanh(pR/2)}\right].
\end{eqnarray} 
For $r_c\gg R$ this propagator exhibits the $1/p^2$- behavior 
both for large and for small  momenta.
\begin{eqnarray} \label{variation}
\Delta(p,  0) &\simeq&  \frac{1}{p^2}\left[1-\frac{2}{pr_c}\right] 
\qquad p\,R\gg 1 \; \textrm{(short distances)}, \nonumber \\
\Delta(p,  0) &\simeq &  \frac{1}{p^2}\left[1 - \frac{R}{r_c}\right]
 \qquad p\,R\ll 1 \; \textrm{(large distances)}.
\end{eqnarray} 
Therefore  the potential is four-dimensional both  for  large and  for small 
distances. Note that the maximal deviation in the coefficient 
of $1/p^2$ is of order $R/r_c$. 

Let us now turn to Minkowski space-time by  rotating $p\to ip_M$.
The propagator takes the form: 
\beq
\Delta(p_M, 0) ~ =~- \frac{1}{p_M^2}\left[\frac{1}{1+ 
(1/p_Mr_c)2\tan(p_MR/2)}\right]~. 
\label{poles}
\eeq
This expression has poles at 
$r_c p_M= -2 \tan(p_MR/2)$. In order to understand better 
this result let us turn to the KK expansion. 
The mode decomposition is given in  Appendix. 
As one would expect,   there is a discrete 
tower of KK states. For $r_c \ll R$ the KK  masses approach  the 
usual  KK spectrum, $m_n=2\pi n/R$. For $r_c \gg R$ all modes 
(except the zero mode) tend to the asymptotes of 
the tangent function: $m_0=0, m_n\simeq 
(2n-1)\pi/R$. This is  in agreement with the pole structure of 
the propagator (\ref{poles}).
Note that the level spacing for fixed $R$ does not change.
For arbitrary $r_c$ the mass of the $n$'th state is in the interval 
$((2n-1)\pi/R, 2n\pi/R) $.

Let us calculate the static potential between two objects on the brane 
which are separated by a distance $r$. For this we perform 
the KK decomposition of the 
corresponding 5D fields. This is done in Appendix. 
The result can be summarized as follows:
\beq \label{potential1}
V(r)~\simeq ~\frac{G_0}{r} + \sum_{n=1}^{\infty} G_n \frac{e^{-m_{n}r}}{r} ~,
\eeq
where for the case 
$r_c\gg R$ we can use  $m_n \simeq (2n-1)\pi/R$ and 
after some simplifications we find (see Appendix): 
\begin{eqnarray}
G_0 & = & \frac{1}{M^3} \frac{1}{r_c+ R} \simeq  \frac{1}{r_c 
M^3}
 \equiv G_\infty \label{zeromode} \\
G_n & = & \frac{1}{r_c M^3} 
\left(\frac{R}{r_c}\right)\frac{2}{[(2n-1)
\pi/2]^2 + R/r_c + (R/r_c)^2}\nonumber \\
& \simeq& \frac{1}{r_c M^3}\left(\frac{2}{\pi^2}\right)\left(\frac{R}
{r_c}\frac{1}{n^2}\right). \label{massive}
\end{eqnarray}
Finally, we derive the potential:
\beq \label{potential11}
V(r)~\simeq ~\frac{G_\infty}{r}  \left(1 + \frac{2}{\pi^2} \frac{R}{r_c}
\sum_{n=1}^{\infty} \frac{e^{-(2n-1)\pi r/R}}{n^2}\right)~.
\eeq
The term in the parenthesis in this expression 
changes slowly with $r$ and the magnitude
of the change is of the order $R/r_c$. 
One could think of (\ref{potential11}) as 
being the Newton potential
with an effective space-dependent Newton's constant. 
To estimate the variation  of the effective gravitational  
constant we evaluate the  
difference of the potentials at two distinct points. 
For a given distance $r$ in the KK sum we can 
neglect the modes with $ n > R/r $ (since these are exponentially 
suppressed)
and replace the exponential by 1 in the remaining terms for which  
$n < R/r$. The variation of the Newton constant for two  
points $r_1$ and $r_2$ 
($r_1 > r_2$) evaluates as follows:
\beq
\frac{G(r_1)-G(r_2)}{G_\infty}~  \sim ~
\frac{R}{r_c}\sum_{n=R/r_1}^{R/r_2} \frac{1}{n^2}~.
\label{difference}
\eeq
Moreover, the potential can be evaluated exactly and 
the expression can be expanded 
for small and large values of $r/R$. For $r/R\ll1$ and  $R\ll r_{c}$,
we find:
\beq\label{potential2}
V(r)~\simeq~\frac{G_\infty}{r}\Bigg[\frac{1}{1+R/r_c} + \frac{R}{r_c}\left(1
 + \frac{4}{\pi} \frac{r}{R}\left(\ln{\frac{r}{R} + C}\right) + O\left
 ((R/r_c)^2, (r/R)^2\right)\right)\Bigg],
 \eeq
where $C=\ln{(\pi/2e)} \simeq -0.54$. Likewise, 
for $r/R\gg1$ and $R\ll r_{c}$, the potential reads as follows:
\beq \label{potential222} 
V(r)~\simeq~\frac{G_\infty}{r}\Bigg[\frac{1}{1+R/r_c} + 
\left(\frac{4}{\pi^2}\frac{R}{r_c} + O\left((R/r_c)^2\right)\right)
e^{-\pi r/R} + \dots\Bigg]~.
\eeq 
The overall relative change in Newton's constant for $r$ in the
interval $[0,\infty]$ is the ratio $R/r_c$. 

\section{Constraints}
\setcounter{equation}{0}

\subsection{Tensor Structure of the Propagator}

In the present model 4D-gravity on the brane is 
mediated by the zero mode graviton with an admixture of the tower 
of massive KK states. The latter
contribution is suppressed both by the parameter $\alpha \equiv R/r_c$ and by the 
KK number $n^2$. 
In order to determine the tensor structure of the graviton 
propagator let us compare 
how it changes when one turns from a massless graviton to 
massive one.
In 4D space the momentum-independent part of the 
tensor structure takes the form:
\beq
{1\over 2}(\eta^{\mu\alpha}\eta^{\nu\beta} + 
\eta^{\mu\beta}\eta^{\nu\alpha}) - 
{1\over 2}\eta^{\mu\nu}\eta^{\alpha\beta}~.
\label{tensor1}
\eeq
This can be seen by using, for instance, the harmonic gauge 
for the fluctuations of the massless gravitational field:
\beq
\partial^\mu h^{(0)}_{\mu\nu}~=~{1\over 2} ~\partial_\nu~h^{(0)\beta}_
\beta~.
\label{gauge1}
\eeq
On the other hand, for a massive graviton one finds:
\beq
{1\over
2}(\eta^{\mu\alpha}\eta^{\nu\beta} + \eta^{\mu\beta}\eta^{\nu\alpha}) - 
{1\over 3}\eta^{\mu\nu}\eta^{\alpha\beta}~.
\label{tensor2} 
\eeq
In the latter case, the equations of motion for a massive graviton
(we use the Pauli-Fierz mass term) 
give  rise to the following relation for the fluctuations:
\beq
\partial^\mu h^{(m)}_{\mu\nu}~=~\partial_\nu~h^{(m)\beta}_
\beta~.
\label{gauge2}
\eeq
Using this condition in the Einstein equations 
naturally leads to the tensor structure given in (\ref {tensor2})\footnote{
Note that the condition (\ref {gauge2}) would be an  inconsistent gauge choice
for the massless gravitons. 
This is in contrast with the vector field case, where
the equation of motion for a massive vector field $A_\mu$ (Proca field)
would require the relation $\partial^\mu A_\mu=0$ and this latter 
is also an acceptable gauge choice for a massless gauge field.}.
Let us now see what happens in the  5D model.
Before compactification one can  choose the harmonic gauge in the 
bulk space:
\beq
\partial^A h_{AB}~=~{1\over 2} ~\partial_B~h^{C}_C~.
\label{gauge5}
\eeq 
This  leads  to the tensor structure of the form (\ref {tensor2}).
Let us consider what happens upon compactification. 
For the massive KK modes, $m\neq 0$, the $\{\mu\nu\}$ components 
of the gauge condition (\ref {gauge5})
will turn into the following expression:
\beq
\partial^\mu h^{(m)}_{\mu\nu}~=~{1\over 2} 
\left (~\partial_\nu~h^{(m)\beta}_\beta~+~\partial_\nu~h^{(m)5}_5 \right )~,
\label{munu}
\eeq
while the $\{55\}$ component takes the form:
\beq
h^{(m)\mu}_{\mu}(x)~=~h^{(m)5}_5(x) ~.
\label{55}
\eeq
Substituting (\ref {55}) into  (\ref {munu}) 
we derive (\ref {gauge2}). Using this, we obtain the 
following expression for the tensor structure of massive KK 
gravitons:
\beq
{\tilde h}^{(n)}_{\mu\nu}(p, 0)~{\tilde T}^{\prime \mu\nu}~\propto~
\nonumber \\
~\left \{  
{\tilde T}_{\mu\nu}{\tilde T}^{\prime \mu\nu} ~-~
{1\over 3}~{\tilde T}^\alpha_\alpha 
{\tilde T}^{\prime \beta}_\beta
\right \}~,
\label{mom1}
\eeq
where ${\tilde T}_{\mu\nu},~{\tilde T}^{\prime \mu\nu}$ are 
Fourier transformed matter source energy-momentum tensors.
Therefore in the approximation $R/r_c\ll1$ the expression for the 
propagator takes the form
(the tensor structure of the zero-mode is that of a 4D theory):
\beqa
G_4^{\mu\nu\alpha\beta}(p)~ \simeq
G_\infty \Bigg(\left ( {1\over
2}(\eta^{\mu\alpha}\eta^{\nu\beta} + \eta^{\mu\beta}\eta^{\nu\alpha}) - 
{1\over
2}\eta^{\mu\nu}\eta^{\alpha\beta}  \right )~{1\over p^2} + \nn \\
 \frac{2}{\pi^2} \frac{R}{r_c}
  \sum_{n=1}^{\infty}{1 \over n^2}
 \left ( {1\over
2}({\tilde \eta}^{\mu\alpha}{\tilde \eta}^{\nu\beta} + {\tilde \eta}^{\mu\beta}{\tilde \eta}^{\nu\alpha}) - 
{1\over
3}{\tilde \eta}^{\mu\nu}{\tilde \eta}^{\alpha\beta} \right )~{1\over p^2 +m_n^2}\Bigg)~ ,
\label{sumG}
\eeqa
where 
\beqa
{\tilde \eta}_{\mu\nu}~\equiv~{\eta}_{\mu\nu}~+~{p_\mu p_\nu \over m_n^2}~.
\eeqa
Thus, the lowest  massive KK modes 
are suppressed  by the ratio  $R/r_c\equiv\alpha$. 
These contributions  might be important. 
Indeed, as shown in \cite{DGP1} 
any ghost-free theory in which massive spin-2 KK 
contribution to 4D gravity is essential should 
suffer from  vDVZ-discontinuity problem \cite{Veltman,Zakharov}.
The reason is that in flat space any massive spin-2 states 
have $5$ polarizations.  Three  of these polarizations 
couple to the conserved stress 
energy tensor with the same strength 
and contribute to the gravitational potential. 
This  induces a finite deviation from the predictions of Einstein 
gravity  for arbitrarily small mass \cite {Veltman,Zakharov}.
Therefore, theories with massive graviton exchange encounter  
phenomenological difficulties and are severely constraint.
We shall try to quantify  the implications of this constraint in our 
model. 

As we discussed above, the massive modes give rise to the distinctive
tensor structure in the propagator  
which alters the predictions for light bending by Sun and 
the precession of  Mercury perihelion.
Therefore, the massive KK modes must be adequately suppressed.
The sum in (\ref {sumG})  is rapidly convergent.  
As a result, the dominant contribution comes from 
a few lightest KK modes.
Using the present experimental data \cite {Will},
this gives the following constraint:
\beq
\alpha\equiv \frac{R}{r_{c}} < 3\times10^{-4}~.
\eeq

\subsection{Planetary Motion}

We shall now examine the constraints on the size of
the extra dimension $R$, coming from  solar system dynamics. 
An object orbiting in the gravitational field will exchange
a certain number of the massive KK modes.  This  is determined by
the distance of the object from the center of the potential.
For  example, Pluto will effectively exchange less KK modes than
Mercury. 
This will be seen as an effective change of Newton's constant and will 
constraint the possible values of $R$. 
In the present  model, the total change in Newton's 
constant felt by an object
falling from infinity towards the center of the potential would be
$R/r_{c}$. From the expansions of the potential (\ref{potential2})
 and (\ref{potential222}) it is obvious that most of the change 
 happens in the range $0<r<R$. 
 
Experimental bounds on the spatial variation of Newton's constant
over planetary distances are quite strong. They come from precise
tracking of planetary orbits with the help of radar beams and
space probes. The results of those measurements are presented in the
work \cite{talmadge} (an updated and  detailed discussion can
be found in \cite{non-newtonian}). 
The modification of the Newton's potential assumed in \cite{talmadge} is
\beq\label{yuk}
\Delta V(r)= \alpha G_\infty \frac{e^{-r/\lambda}}{r},
\eeq
where $\alpha$ and $\lambda$ are parameters that describe deviations
from standard potential. A nonzero value of these parameters would result
in  deviations from the Kepler's third law and would induce  anomalous 
 precessions  of  the planets' perihelia. The
precise measurements of orbits of the inner planets  
and satellites put bounds on the values of $\alpha$ and $\lambda$ that
are summarized in  Ref. \cite{talmadge,non-newtonian}.

Let us  
relate approximately the 
parameters of our model with the parameters in (\ref{yuk}).
For  simplicity we will use the following estimate
\beqa\label{estim}
 \Delta V(r) &\simeq&
\frac{G_\infty}{r}  \left( \frac{2}{\pi^2} \frac{R}{r_c}
  \sum_{n=1}^{\infty} \frac{e^{-(2n-1)\pi r/R}}{n^2}\right) 
  =\frac{2R}{\pi^2r_c} G_\infty \frac{e^{-r\pi/R}}{r} +
  \frac{2}{\pi^2} \frac{R}{r_c}
  \sum_{n=2}^{\infty} \frac{e^{-(2n-1)\pi r/R}}{n^2}
    \nonumber \\
&<  & \frac{2R}{\pi^2 r_c} G_\infty \frac{e^{-\pi r/R}}{r} \left(1 +
 \sum_{n=2}^{\infty} \frac{1}{n^2}\right)=  \frac{R}{3r_c}G_\infty 
 \frac{e^{-\pi r/R}}{r}.   
\eeqa
With this estimate we can relate the parameters of our model
with the parameters in (\ref{yuk})
\beq\label{parest}
\alpha\sim\frac{R}{r_{c}},\quad \lambda\sim R.
\eeq

The bound on $\alpha$
is very stringent ($\alpha\leq 10^{-9}$, \cite{talmadge,non-newtonian})
for the values of $\lambda$ in the range from the Earth radius to
the Solar System size. If we assume $M\sim $TeV we conclude that the size of
extra dimension $R$ must be either smaller than the Earth radius, or 
bigger than the solar system size.

Let us discuss these two possibilities
separately. We start  with the case when $R$ is 
larger then the solar system size.
In this case, the estimate (\ref{parest}) is actually not the most precise one. The  bounds
on $\alpha$ and $\lambda$ shown in \cite{talmadge}  were derived using the 
spatial variation of Newton's constant that appears in the force, rather than in 
the  potential.
 We can  compare the correction to the force in our model
and the same correction  for the pure Yukawa type modification
, both shown in Fig.1 .
We see that the force correction
in our model is a  much steeper function on small values of $r/R$.

The measured quantity is the parameter $\eta_{p}$ that describes
deviations from third Kepler's law. The ratio of semi-major axis can
be measured $(a_{p}^{meas})$ and compared to prediction of Kepler's law
$(a_{p})$ . The following  equation relates $\eta$ and the value of Newton's
constant at the Earth ($G_{E}$) and at the planet's orbit ($G_{p}$). 
\beq
\frac{a_p}{a_p^{meas}}= 1 + \eta_p = 
\left(\frac{G_p M_s}{G_EM_s}\right)^{1/3}.
\eeq
Newton's constant is the one that appears in the force law, and 
$M_s$ is the mass of the Sun.
The bound on the parameters $\eta_{p}$ is of the order $10^{-9}-10^{-10}$ for the
inner planets.

\vspace{5mm}
\centerline{\epsfig{file=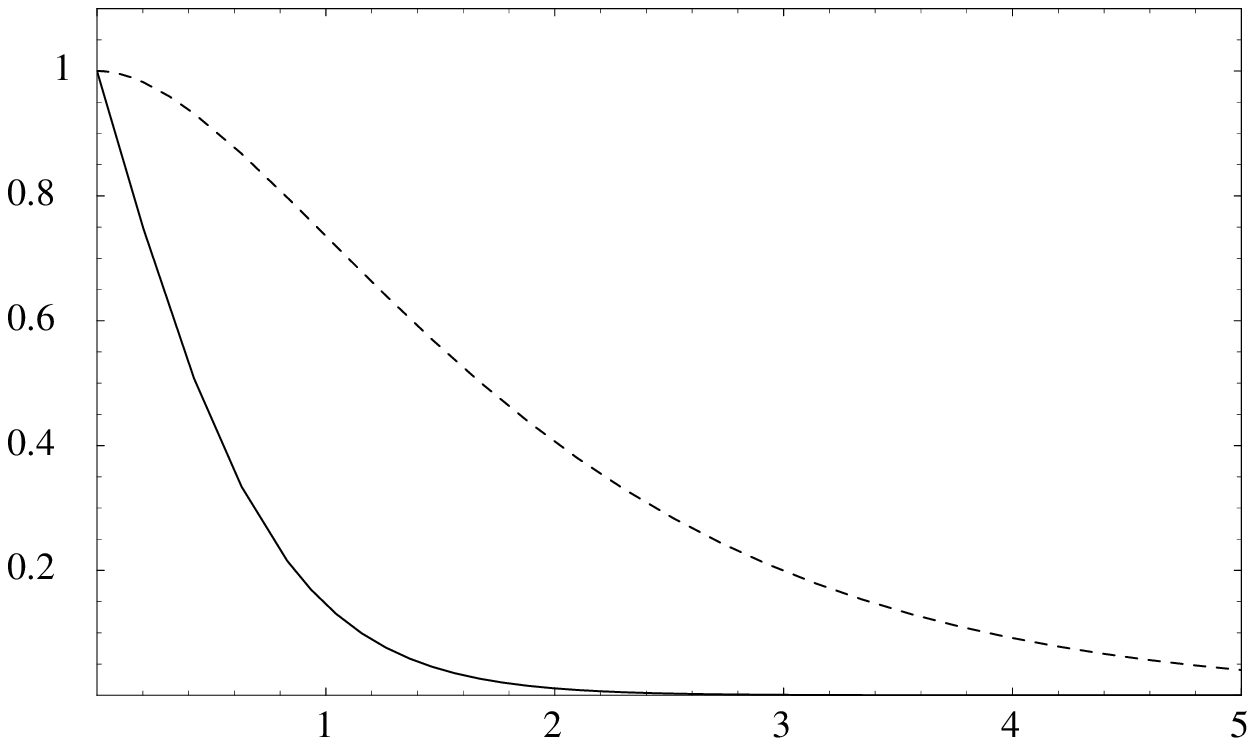,width=10cm}}
{\footnotesize\textbf{Figure 1:} The correction to Newton's force for our model 
(solid line, $R/r_{c}=10^{-4}$)
and for the Yukawa type correction (dashed line, $\alpha=10^{-4}$). 
The relative correction  (normalized to the strength of Newton's force) is given
is units of $10^{-4}$. The $x$ axis measures $r/R$ for our model and 
$r/\lambda$ for the Yukawa correction. }
\vspace{5mm}

The most precise bound on the spatial variation of Newton's constant $(\eta_p)$
comes from the  relation between the Earth and inner planets, i.e., in the regime
of small $r/R$, for $R$ greater then the solar system size. 
In that regime, the  change in force will be bigger for our model than
for the Yukawa type correction and we have to compensate that effect
by choosing somewhat larger values of $R$ and $r_{c}$. Numerical analysis
shows that with the choice of $R\approx 10^{16}$m, and $R/r_{c}\approx
10^{-4}$, the bounds can be safely satisfied. Since the Planck scale is
roughly given by $r_{c}M^3$, an  increase of $r_{c}$ by some factor
would lower the fundamental Planck scale for a much milder value
of a cubic root. With the new bound of $R>10^{16}$m
and $R/r_{c}>10^{-4}$ the fundamental Planck scale is pushed to 
approximately $100$ GeV.  

Similarly, we calculated  the contribution to the anomalous precession 
of Mercury and Mars in our model,  and we found  that it can be 
within the accepted values  if the above limits 
on $R$ and $R/r_c$ are satisfied

The second   possibility 
is to take the size of the extra dimension to be anywhere   less than the 
Earth radius ($\sim 10^7$ m),  in order 
to satisfy  constraints from Earth-Moon and artificial  satellite
experiments. This ``regime'' is much safer and the bounds are very weak,
as can be seen in \cite{talmadge,non-newtonian}. 
In the  case $R\sim 10^7 $ m, $r_c$ can range 
form $\sim 10^{11}$ m  to $\sim 10^{19}$ m, 
corresponding to a fundamental Plank scale of 
$\sim 10$ TeV down to $100$ GeV, respectively. 
 Of course the value of $R$ can be lowered  at will,
down to $1/M$, while keeping $r_c$ fixed in order to generate the correct 
value for the four dimensional Plank scale from
 the 'low' fundamental scale $\sim 1$ TeV  .      
To point out 
the importance of our result in this 
regime, we stress that our model makes possible the existence of a single  
extra dimension of very large  size without contradiction with present 
observations, while keeping the value of the fundamental Plank scale 
around the TeV.

The mechanism of 
suppression of heavy modes  in our framework can relax the high energy 
constrains on the  models  
with  submillimeter extra dimensions.

To summarize, there are two windows of the parameter space 
in the present model. 
The first one, in which the extra dimension has astronomic 
scale , is strongly constrained  by the solar system observations  which 
force $R \geq 10^{16}$ m (for $R/r_c \simeq 10^{-4}$). 
In the second scenario $R$ can be anywhere below the  Earth radius and $1/M$.

\subsection{Star Cooling}

The most severe constraint on theories with low scale gravity comes from 
astrophysics  since a  stellar object can cool by radiating bulk 
gravitons\cite{ADD2}.
Let us estimate the rate of star cooling in the present 
case. Let the temperature inside the star be $T$. Then only the gravitons 
with masses $m_n < T$ are produced efficiently. The rate of emission of the  $n$th KK 
graviton  is given by
\beq
 \Gamma_n \sim T^3G_n~.
\eeq
We have to sum  over all states up to $n \sim TR$. 
This gives
\beq
\Gamma_{total}(T) \sim {2T^3~R \over \pi^2 ~M^3~r_c^2} ~\sum_1^{RT}~
 {1\over n^2}
~\simeq~
{T^3R \over 3M^3r_c^2}~ \simeq {\alpha\over 3} {T^3 \over M_P^2}.
\label{coolrate}
\eeq 
This is less than the rate of just zero 
mode graviton production, and is totally negligible. 
 Note that the contribution coming from the
emission of a pair of KK gravitons via the virtual zero mode graviton
exchange is of the same order, and is practically insensitive to the value
of $R$ and $M$. To see this let us consider a scattering process on the
brane with a virtual zero mode graviton emission which subsequently
decays into a pair of KK gravitons.
The example of such a process can be a gravi-brehmstrahlung process,
an electron scattering in the field of a nucleus and radiating gravitons. Let
the typical energy in the process be $E$. For definiteness we take $E >
M$, in order to put the strongest constraint. The rate of the process then
becomes 
\begin{equation}
 \Gamma \sim {E^3 M^2\over M_P^4}(MR) \sim  \alpha{E^3 \over M_P^2}~,
\end{equation}
where we have cut-off graviton momenta by $M$. Surprisingly,
this rate is similar to the one of single KK production, due to the
suppression of direct KK production rate. Most importantly it has no
dependence on either
$R$ or $M$. Thus the KK emission in high-energy processes {\it a priory}
places no bound on either of this parameters. However, we do not expect
$M$ to be smaller than the inverse mm scale, due to no observed deviation
from Newtonian gravity on larger distances, which gives the upper limit on
the size of extra dimension $R \sim \alpha 10^{63}$cm.
Thus the star cooling 
places no constraint on our scenario. It is useful to contrast this with 
the scenario of\cite{ADD}. There, the dominant contribution to the cooling 
process comes from the production of heaviest modes, due to their high 
number. In the present case however, the wave functions of the heavy modes are 
suppressed on the brane and therefore cannot be produced efficiently. This 
is a generic property of the given framework with the induced kinetic term:
in a high-energy process on 
the brane, the lightest KK states are the ones produced most efficiently. 
The extra dimension is more transparent for the softer modes!

\subsection{Cosmology}

In this section we will consider cosmological constraints coming from
the overproduction of bulk states.
Not surprisingly, in analogy with star cooling, the 
cosmological constraints turn out to be rather mild.
In order to be as model independent as possible,
we shall discuss the following initial
conditions for the hot big bang:
 
 1) The bulk is virtually empty;

2) The brane states are in thermal equilibrium at some
temperature $T_{brane}$,  which can not be higher than the
``normalcy'' temperature 
$T_*$. 

The  normalcy temperature 
is defined as  the temperature below which
the Universe expands as being effectively four-dimensional.
This requirement then automatically  restricts $T_{brane} < T_*$.
Obviously, we would like $T_*$ to come out at least as high as the
nucleosynthesis temperature. As we shall see below, $T_*$ could 
be much higher, even as high as the cut-off scale ($T_* \sim M$). 
This fact
allows for the standard nucleosynthesis scenario
to proceed unaffected, and also to implement one of the 
conventional baryogenesis mechanism, for generating the baryon
asymmetry of the Universe.

As discussed in\cite{ADD2}, in theories with large extra dimensions
the overproduction of  bulk KK states can alter the standard cosmological
expansion in two different ways. First, the energy density on the brane 
changes due to the ``evaporation'' into the bulk states. Secondly, the
produced bulk states may dominate 
the energy density and over-close the universe.
Let us consider constraints coming from these two effects separately.

\vspace{0.1in}

{\it Cooling by Evaporation into Bulk States}

\vspace{0.1in}

 As estimated above for star cooling at temperature $T$ the evaporation
rate into the bulk gravitons is given by (\ref{coolrate}). The resulting
change
of the matter energy density on the brane
due to evaporation is given by
\beq
 {d\rho\over dt}|_{evaporation} \sim - T^4
\Gamma_{total}(T) 
\sim - {T^7 \over M_P^2}\alpha~.
\label{coolrateu}
\eeq
Note that this is by a  factor $\alpha$ smaller than the cooling rate due to
the production of the standard zero-mode graviton, and is totally negligible
in comparison to the cooling rate caused by the cosmological expansion
\beq
 {d\rho\over dt}|_{expansion} \sim
-3H\rho\sim -3{T^2\over M_P}\rho~,
\label{coolexpansion}
\eeq
where $H$ is the  Hubble parameter.
For instance, in the radiation dominated epoch ($H \sim T^2/M_P$),
the ratio of the two rates is
\beq
 {{d\rho\over dt}|_{evaporation} \over  {d\rho\over dt}|_{expansion}}
\sim {T\over M_P} \alpha. 
\eeq
This  is a very small number even for $T\sim$ TeV.
The reason for this suppression can again be understood from the
'infrared transparency' of the theory:
since the heavier
gravitons' wave-functions are  strongly  suppressed  on the brane,
their production is not efficient enough to affect
the brane-cooling process.

\vspace{0.1in}

{\it  Overclosure by Gravitons}

\vspace{0.1in}

Due to their  suppressed couplings, KK gravitons 
are harder to produce on the  brane. 
Moreover, due to the same suppression
the KK modes are more long-lived 
as well. Thus, we have to examine constraints coming from their
possible overproduction.

The lifetime of a ``cold'' KK graviton of mass $m$ 
can be written as follows:
\beq
\tau_m \sim {M^2_P (r_c R) \over m}~.
\eeq
This  turns into the following relation upon using (\ref{relation})
\beq
\tau_m \sim \left({M_P \over M}\right)^6 {\alpha \over m}~.
\eeq
Thus, even the gravitons as heavy as $m\sim M$ are stable for all the
practical purposes and can over-close the universe if produced
with a sufficiently large number density. This puts some bound on
$T_*$ which we shall estimate below. Assuming that all the produced
KK gravitons are stable and do not decay back to  the brane,
the energy density which is ``pumped'' into the bulk
due to the brane-evaporation (in the process of the normal expansion)
is
\beq
\delta\rho = \int_{t_{in}}^{t{eq}}{d\rho\over dt}|_{evaporation} dt
\sim \int_{T_{eq}}^{{T_*}}  
{T^4 \over M_P}\alpha dT \sim {T_*^5 \over M_P}\alpha~.
\label{pumpedrho}
\eeq
Here $T_{eq}\sim $eV, is the standard crossover temperature
and we neglected
the later period of matter domination as well
as the order one factors in the integral.
Most of the energy density gets pumped in the bulk during one Hubble
time after $t=t_{in}$. Thus, soon after the initial time the 
bulk becomes
populated by KK states with the energy density given 
by (\ref{pumpedrho}).
Most of these particles are relativistic and their energy density
will redshift as radiation ($\rho \sim T^4$).
However, to make our bound most conservative let us assume that 
most of the energy density is stored in the cold KK modes that
redshift as matter
($\rho \sim T^3$). Even in this case  their energy density at the time
of standard crossover will be given by
\beq
\rho_{KK} \sim {T_*^2 \over M_P}\alpha T_{eq}^3~.
\label{finalrho}
\eeq
Requiring that this be much smaller than the energy stored in the usual
matter (at the same time $t = t_{eq}$), which, by definition, is 
$\rho_{matter} \sim T_{eq}^4 \sim (10 \textrm{eV})^4$,
we get the following bound on $T_*$
\beq
T_*^2 <  {M_P\over \alpha} eV = {10^{10} \over \alpha}{\rm GeV}^2~.
\label{finalrho1}
\eeq
This bound can be easily satisfied even for $T_* \sim M$.

In conclusion, we see that in contrast with the scenario of
\cite{ADD} there are no essential cosmological constraints 
from overproduction of bulk states due to the brane cooling.
 We have to stress, however, that this analysis can not capture more model
dependent possibilities. For instance, if the KK states are produced by
some other means in the 
early universe, they could either over-close the Universe or
serve as unusual (and interesting) dark matter candidates.
To avoid the overproduction we have to assume 
that there was a period of the inflation that diluted the bulk and reheated 
only the brane \footnote{The
inflationary solutions in the limit of $\alpha = \infty$ were studied
in\cite{cedric}.}.

\section{On the Origin of the Crossover Scale}

In this section we shall discuss a possible origin of the large
distance scale $r_c$. 
The goal is to explain the large 
coefficient in front of the four-dimensional curvature term
in Eq. (\ref{4DR}).
As already noted, in the  effective field theory picture this term is not
constrained by any symmetry and can emerge  with an  {\it a priory}
unrestricted coefficient. However, it is desirable to have better
understanding of this issue. As suggested in \cite {DGPind,DG},
this term is induced due to quantum loops of the
states which are localized on the brane and 
which interact  with high-dimensional
gravity. The resulting strength  depends  on the 
number of such states as well as their masses \cite {DG}. 
Thus, the large mass hierarchy could be obtained due to the 
large multiplicity of states propagating in matter loops on the 
brane.  However, there can be other effects
that may significantly contribute to the magnitude of this term. 
One possibility is
to consider a brane-scalar field $\xi$ which is 
non-minimally coupled to gravity
\beq
S_{\rm brane}~= M^2~\int d^{4}x~
\sqrt{|{\rm det}{\bar g}|}~ {\rm exp}\left ( {\xi\over M} \right )
{\R4}(x)~.
\label{xi}
\eeq
One could  assume that, due to some dynamics, $\xi$ develops a
vacuum expectation value close to the scale $M$. 
This is in no contradiction
with any fundamental principle, as long  as $\xi$
is somewhat lighter than $M$.
Shifting the field $\xi\rightarrow <\xi> +  \xi$ we end up with an
exponentially large scale $r_c = M^{-1}{\rm e^{{<\xi> \over M}}}$. 
Note that
in the shifted vacuum the  theory  is consistent  
up to the energies of order $\sim M$. Moreover, 
the perturbative treatment  is valid  since the emission of $\xi$
quanta are suppressed by the powers of $M^{-1}$.

Although at a  first glance the fact
that the VEV is larger than the cut-off of the theory might seem a 
bit unnatural, we should  stress that there
is nothing unusual in this fact. There exist  many well defined examples
both in string theory as well as in KK theories when this is the case.
After all, the solution of the hierarchy problem in
\cite {ADD} can also be understood in this way.
Indeed, the large value of the Planck scale is generated by
the size of the extra dimension, or equivalently,
by the VEV of the four-dimensional
scalar field, the radion, which gets expectation value exceeding by many orders of
magnitude the cut-off of the theory.

Another well known example of this kind  can be found in $D$-branes. 
It is well known that in the BPS limit
the separation of two parallel $D$ branes
can be understood as the Higgs effect in  
the brane worldvolume gauge theory. 
The expectation value of the
canonically normalized Higgs field is related to the string scale,
$M_S$, as
\beq
<\xi> ~\sim ~rM_S^2~,
\eeq
where $r$ is the inter-brane separation. This latter  
can be much larger than $1/M_S$.
In fact the limit $r >> M_S^{-1}$ is well defined perturbatively
and corresponds
to the infrared limit for the bulk gravity. 
Although the world volume scalar
field acquires an  expectation value which is much larger than 
the fundamental scale,
the theory is well defined. The two-brane system becomes populated
with  states of mass $\sim  rM_S^2$ that are much heavier than
the string scale
(corresponding to the stretched string modes in the original theory).
It may not be impossible that in a suitable framework like this,
the brane separation can set the large coefficient of the brane-induced
curvature term.

\section{Implications  for Sub-millimeter Dimensions}

So far we have been  studying  the domain   $R < r_c$.  As we
have shown, in such a case $R$ can be of astronomical size.
In fact, it can be arbitrarily large (if $r_c$ is taken to be large).
In this section, we shall study the opposite regime $R > r_c$. The case
$R \rightarrow \infty$ with one extra dimension was already discussed
in \cite{DGPind}. In this case the massive KK states
which are lighter than $r_c$ have unsuppressed 
wave-functions on the brane. Thus, they 
couple to the brane-matter  with  a  strength comparable to that
of the zero-mode graviton. Therefore, the masses of the
KK modes must be large enough in order to avoid unacceptable
deviations  from Einstein's  gravity. Precision measurements of
light-bending  and the precession of Mercury perihelion  exclude the range of 
masses at the inverse astronomical scale. 

The shorter ranges are constrained by
the precision measurements of $G$ discussed above. These constraints
push $R$ to be somewhere around the inverse millimeter range, in fact,
according to recent measurements \cite{adelberger} $M^{-1} > 250 \mu$ or so.
In such a regime,   planetary dynamics  is  insensitive to
extra dimensions, just like in the original scenario of \cite{ADD}. 
However, the other predictions of this framework are dramatically modified.
The situation is somewhat peculiar for $M^{-1}\ll r_c \ll R$. On the one hand,
the predictions of \cite{ADD} for  table-top
gravitational experiments \cite{tabletop} are unaffected.
On the other hand all the current
cosmological and astrophysical constraints are lifted, and the collider
signatures are dramatically modified. This happens  
due to the suppression of heavy KK production at high energy colliders.

In order to explicitly demonstrate this 
let us consider the star cooling process. 
This process puts the
most stringent constraint on the scenario
of \cite{ADD} for the sub-millimeter dimensions. 
For a given temperature $T$
the production rate of
$n$th KK state inside the star is suppressed as
\beq  
\Gamma \sim {2T^3 \over M_P^2 } \sum_1^{RT}
 \frac{1}{1+ n^2 (\pi r_c/R)^2}~.
\eeq
Assuming that $r_c \gg 1/T$, we can evaluate the contribution of 
the KK modes with $m < 1/r_c$ and  $m > 1/r_c$ 
separately and show that  these are both of order
${T^3R\over M_P^2r_c}$. Using the relation $M_P^3 \sim M^3R$ we can bring
the total contribution to the form
\beq
 \Gamma \sim {T^4 \over M^3(Tr_c)}~.
\eeq
Notice that there is an extra suppression factor $\sim (Tr_c)$
with respect to the standard case of \cite{ADD}. This indicates that
for sufficiently large $r_c$ all the bounds can be avoided.
Suppression of the bulk graviton production in other high energy processes
can be analyzed in a similar way. The peculiarities of the spectrum of the 
model indicate  that many of the
experimental constraints on theories with  large extra dimensions 
must be reconsidered in the light of the present discussion.

\section{Conclusions}

In this paper we studied the scenario of Ref. \cite{DGPind}
with graviton kinetic term on the brane with one extra compact dimension.
We showed  that the
existence of the graviton kinetic term on the brane
allows for a  novel framework with the high-dimensional
fundamental Planck scale $M\lesssim$ TeV 
(while the particle physics scale is bigger than a ${\rm TeV}$), 
and a single {\it flat} extra dimension with
the size which can be either smaller than $10^7$ m or 
larger than $10^{16}$ m. 

The crucial role in generating usual 4D Einstein gravity on the brane
is played by the brane-induced graviton kinetic term of 
reference \cite{DGPind}.
The strength of this term is governed by a distance scale $r_c$
which together with the size of the extra dimension defines 
the value of the four-dimensional Planck mass (see (\ref{relation})).

There are two phenomenologically interesting regimes. 
The first one is achieved when  $R \ll r_c \ge 10^{16}$ m.
In this case the four-dimensional gravity on the  brane
is mediated by a single {\it non-localized} zero mode graviton, 
both at large and at short distances. 
The mass spacing of the KK modes is similar
to that of  an ordinary flat compact dimension.  
However, the wave-functions of heavy KK modes
are suppressed on the brane by the ratio $R/r_c$. 
This  gives rise to the effect of 
``infrared transparency'' \cite{DG} of the extra space.
We studied  the constraints imposed by 
precision gravitational measurements at all scales, as well
as restrictions due to various astrophysical and cosmological effects. 
We found that the model is compatible with all those data.
A crucial experimental test of this scenario  could arise by  
observing  deviations from Newtonian and Einstein's 
gravity practically at any scale. This includes  precision 
studies  of both relativistic and non-relativistic effects.
In the present framework 
the states that live in the bulk can be a rather
exotic source of dark matter. 
At distances  $r\gg R$  their interaction is that of an ordinary
dark matter. However, for  $r\ll R$ they interact via much stronger
gravitational potential which scales as $1/r^2$. 
Furthermore,  these states interact with the observable
matter by much weaker gravitational force. 

Another interesting
limit is $R > r_c$ and $N\ge 2$ (in $D=4+N$ dimensional space). 
In this case  the compactification radius 
$R$ is constrained to be in a sub-millimeter domain. 
Although this seems to be similar to the scenario
of \cite{ADD}, nevertheless, 
the framework is modified dramatically due to the graviton kinetic
term on the brane. For instance, if 
$M^{-1} < r_c < R$, the table-top predictions of scenario \cite{ADD}
are unaffected and one should still expect deviations from 
Newton's law at scales $r\sim R$. 
However, production of heavy KK gravitons
is strongly suppressed. This lifts all the high-energy 
constraints. We have explicitly demonstrated this fact by reevaluating
the constraints coming from the process of star cooling.
Our analysis demonstrates the crucial importance of the brane-induced
graviton kinetic term for the phenomenological studies.

\vspace{0.4cm}

{\bf Note added}
\vspace{0.1cm} \\

After this work was completed, the related work \cite {Lykken}
appeared, which  discusses the role of brane-induced kinetic term 
on compact dimensions in an example of a scalar field theory.

\vspace{0.4cm}

{\bf Acknowledgments}
\vspace{0.1cm} \\

The work of GD was supported in part by David and Lucille Packard
Foundation Fellowship for Science and Engineering, by Alfred P. Sloan
foundation fellowship and by NSF grant PHY-0070787. 
 The work of GG  is supported by  DOE Grant DE-FG02-94ER408.
The work of MK was
supported in part by David and Lucille Packard Foundation.
FN is supported by the NYU McCraken Fellowship.

\vspace{0.2in}

\section{Appendix}
\setcounter{equation}{0}

In this Appendix we derive the expression for the force mediated by a  
scalar field in five-dimensional space-time with one compact dimension and  
an induced kinetic term on a 3-brane. 
We denote the coordinates $x^{A}\equiv(x^{\mu},y),y\in[0,R].$
Our starting point is the following five-dimensional Lagrangian
\beq \label{L5}
{\mathcal L}=  M^3 \partial^A \Phi \partial_A \Phi + {\M4}^2
 \delta(y) \partial^\mu
 \Phi\partial_\mu \Phi,
\eeq
where $\Phi(x^A)$ has dimension 0,  $M$ is the fundamental five-
dimensional mass scale, and  $\bar{M}^2$
is the scale which is dynamically generated by 
the interaction with matter that lives on the brane \cite{DGPind} .  
If we absorb the factor  $M^{3/2}$ into  the 
redefinition of the field then $\Phi$ becomes a canonically 
normalized 5D scalar. The  ratio 
$r_c={\M4}^2/M^3$ appears in the field equation as follows:
\beq \label{field eq1}
\partial^A \partial_A \Phi + r_c \delta(y) \partial^\mu \partial_\mu \Phi 
=0.
\eeq
Let us decompose the field  $\Phi$ into the following modes
$\Phi=\sum_{m} \phi_n(y)\sigma_n(x^\mu)$.   $\sigma_{n}(x^{\mu})$
satisfy  the four dimensional Klein-Gordon equation
$(\partial^\mu \partial_\mu \, +\, m_n^2)\sigma_n =0$, where $m_n$'s 
are to be  determined. 
The functions $\phi_n(y)$ set the  profiles of the field in the fifth 
dimension. From (\ref{field eq1}) (with the use of $\partial^A \partial_A =
\partial^\mu \partial_\mu -
\partial_y^2$) we get the ``Schr\"oedinger equation'' for $\phi(y)$
\beq\label{field eq2}
\bigg( \partial_y^2 + m^2 + r_c m^2 \delta(y) \bigg) \phi(y) =0.
\eeq
This equation is to   be considered on a circle of length $R$ with the
periodicity condition $\phi(y + R)=\phi(y)$. The problem is equivalent
to that of a wave-equation for an infinite space with an array of 
delta-function type 
potentials  located at 
$y=nR , n=-0, \pm 1, \pm 2, \ldots$, with the identification 
$y\equiv y+R$. The procedure for finding  the solutions 
is standard: We solve the equation in two
neighboring regions $ I\equiv[-R, 0]$, $II\equiv[0, R]$, match the 
wave-function 
and its derivative at the boundary, and then identify the two regions by 
imposing  the periodicity condition $\phi_I(y-R)=\phi_{II}(y)$. The
solutions in each region are 
\begin{eqnarray}
&\left.I\right)& \qquad   \phi_I(y)= A e^{imy} + B e^{-imy} \nonumber \\
&\left.II\right)& \qquad   \phi_{II}(y)= C e^{imy} + D  e^{-imy},
\end{eqnarray}
where the coefficients $A, B, C, D$ are to be determined by the following 
conditions: (a) periodicity; (b) continuity at $y=0$; (c) matching of 
first derivatives at $y=0$:
\begin{eqnarray}
&\left.\textrm{a}\right)   &   A e^{im(y-R)} +B e^{-im(y-R)}=C e^{imy} 
+ D  e^{-imy} \nonumber \\  
&\left.\textrm{b}\right)   &  A+B=C+D  \\
&\left. \textrm{c}\right) &  C-D  - A (1+ir_c m) + B (1- i r_c m) =0.
 \nonumber
\end{eqnarray}
Solving these  algebraic equations we determine the coefficients $B, C, D $ 
in terms of $A$. This latter  can in turn be found 
from the normalization condition on $\phi$  
\beq
D~=~A ~\qquad \qquad B ~= ~C ~= ~A ~e^{-imR}~,
\eeq
amended by the quantization condition for the masses
\beq \label{mass}
-\frac{r_c}{R}\left(\frac{mR}{2}\right) = \tan \left(\frac{mR}{2}\right).
\eeq
The nonlinear equation (\ref{mass}) is solved by graphical method as  shown
in  Fig.2

\vspace{5mm}
\centerline{\epsfig{file=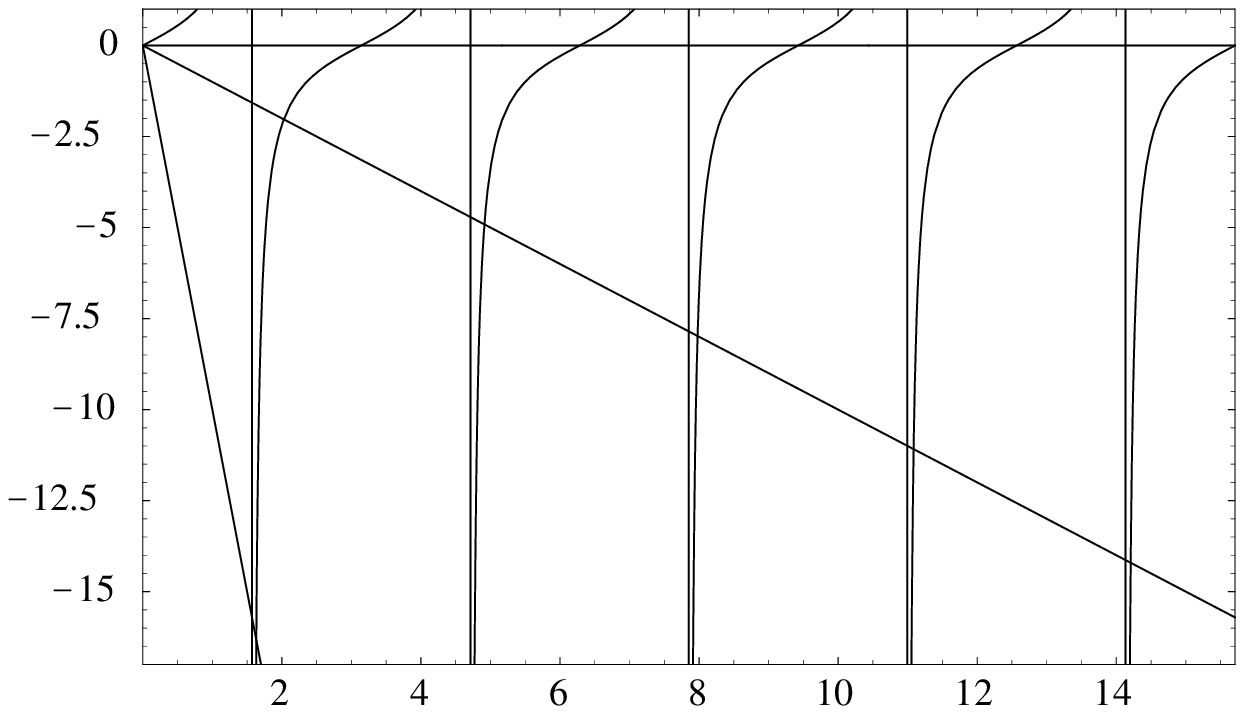,width=10cm}}
{\footnotesize\textbf{Figure 2:} 
 Nonlinear condition for the mass spectrum (\ref{mass}). The $x$ axis is given
in units of $mR/2$. The functions shown are $\tan$ and the lines with slopes
$-r_{c}/R=-10,-1,0$. For $r_{c}/R=0$ the spectrum is standard Kaluza Klein
spectrum $m_{n}=2n\pi/R$, while for large $r_{c}/R$ it is given by 
$m_n\simeq (2n-1)\pi/R$. }
\vspace{5mm}

 We see that for $r_c \ll R$ the masses approach the usual 
Kaluza-Klein  spectrum, $m_n=2\pi n/R$. For $r_c \gg R$ all modes (except 
zero mode) approach the asymptotes of the tangens: $m_0=0, m_n\simeq 
(2n-1)\pi/R$.
The level spacing  for fixed $R$ does not  change.
For arbitrary $r_c$ the mass of the $n$'th state  is 
in the interval $((2n-1)\pi/R, 2n\pi/R) $.

Let us now study  the behavior of profiles in the fifth dimension. 
First we have to normalize $\phi_n(y)$ to unity. 
This fixes the coefficient $A_{n}$ 
\begin{eqnarray}\label{norm}
A_n &=&  \frac{1} {\sqrt{2R}}\frac{1} {\sqrt{1-\frac{r_c/R}{1+r_c^2 
m_{n}^2/4}}}
 \qquad  n\neq 0, \nonumber \\
A_0 &=& \frac{1} {2\sqrt{R}}.
\end{eqnarray}
For $r_{c}\gg R$ the coefficient $A$  
depends very weakly  on the mass of 
the state and can be  approximated by  $1/\sqrt{R}$ for all the modes.
In Fig. 3 we show the modulus squared of the wave-function for the zero 
mode (constant) and the lowest three modes for the choice $r_{c}/R=10$.

\vspace{5mm}
\centerline{\epsfig{file=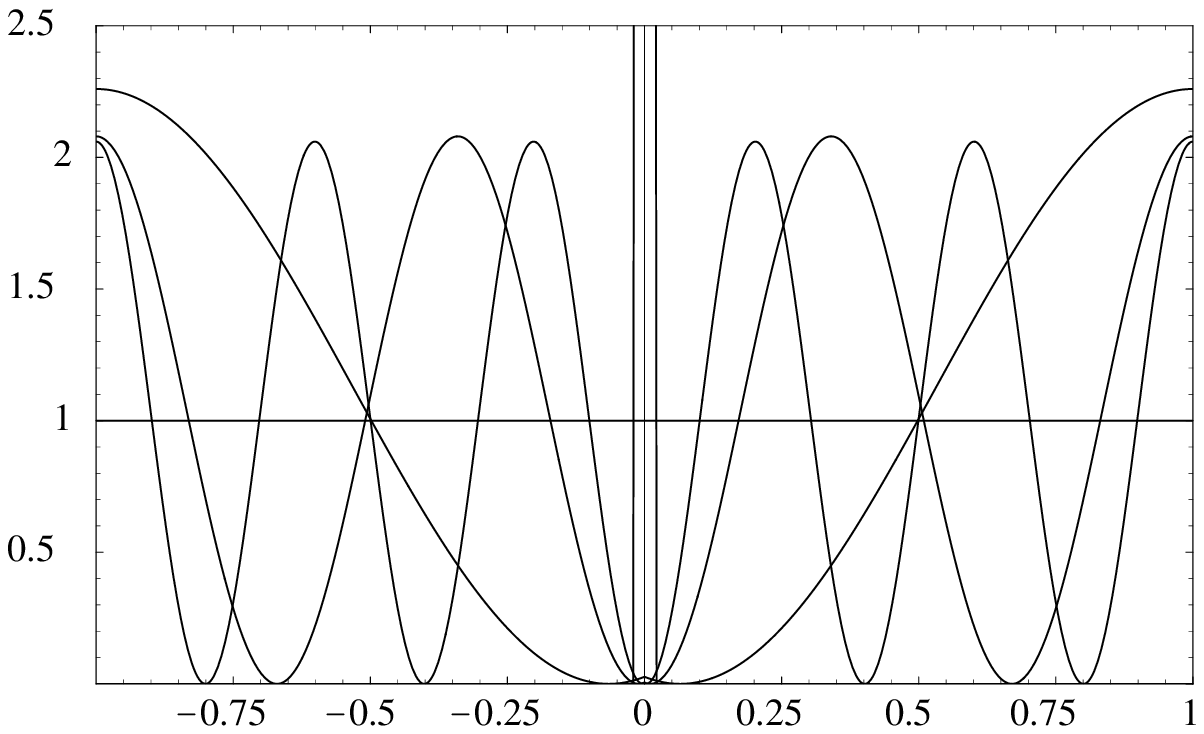,width=10cm}}
{\footnotesize\textbf{Figure 3:} Moduli squared of the wave functions for the zero mode and
three lowest modes. The fifth coordinate is shown on the x axis and the units
are $R/2$ (i.e. the whole range of the extra dimension is covered). The modulus 
squared is plotted on the y axis and the units are $1/R$ (i.e. the modulus of 
the zero mode
is equal to one). The dependence is shown for the zero mode (constant), and 
lowest three modes for $r_{c}/R=10$. The position of the brane at 
the origin is figuratively  sketched as the thin ``barrier''.}
\vspace{5mm}

The quantity which  determines the  coupling of the KK 
modes to the brane matter is the modulus squared of 
the wave-function at  $y=0$:
\begin{eqnarray} \label{suppressed}
\left|\phi_{n}(0)\right|^2  &=& \left|A + B\right|^2 =\left|A 
\right|^2\left|
 1 + e^{-im_{n}R}\right|^2 \nonumber \\
&=& \left|A \right|^2 \frac{4}{1+ \tan^2\left(m_{n}R/2\right)} 
=\left|A \right|^2 \frac{4}{1+ r^2_c m_{n}^2/4 }.
\end{eqnarray}
Here, in the last line
we used the condition (\ref{mass}). 
From (\ref{suppressed}) we deduce that higher KK modes are suppressed
on the  brane compared to  the zero-mode. 
The suppression factor is exactly  the  same 
as in theories with infinite volume extra dimension \cite{DGP1}. 
The mass of the $n$'th mode is of order $2\pi n/R$, so we can rewrite 
(\ref{suppressed}) in terms of our initial parameters $R, r_c$
\beq 
\frac{\left|\phi_n(0)\right|^2}{\left|\phi_0(0)\right|^2} \simeq 
 \frac{1}{1+ n^2 (\pi r_c/R)^2} .
\eeq
For $r_c \gg R$ this  suppression is substantial even for the
lowest  massive KK states. In Fig. 4 we show the modulus squared of the three
lowest massive modes near the origin for $r_{c}/R=10$.

\vspace{5mm}
\centerline{\epsfig{file=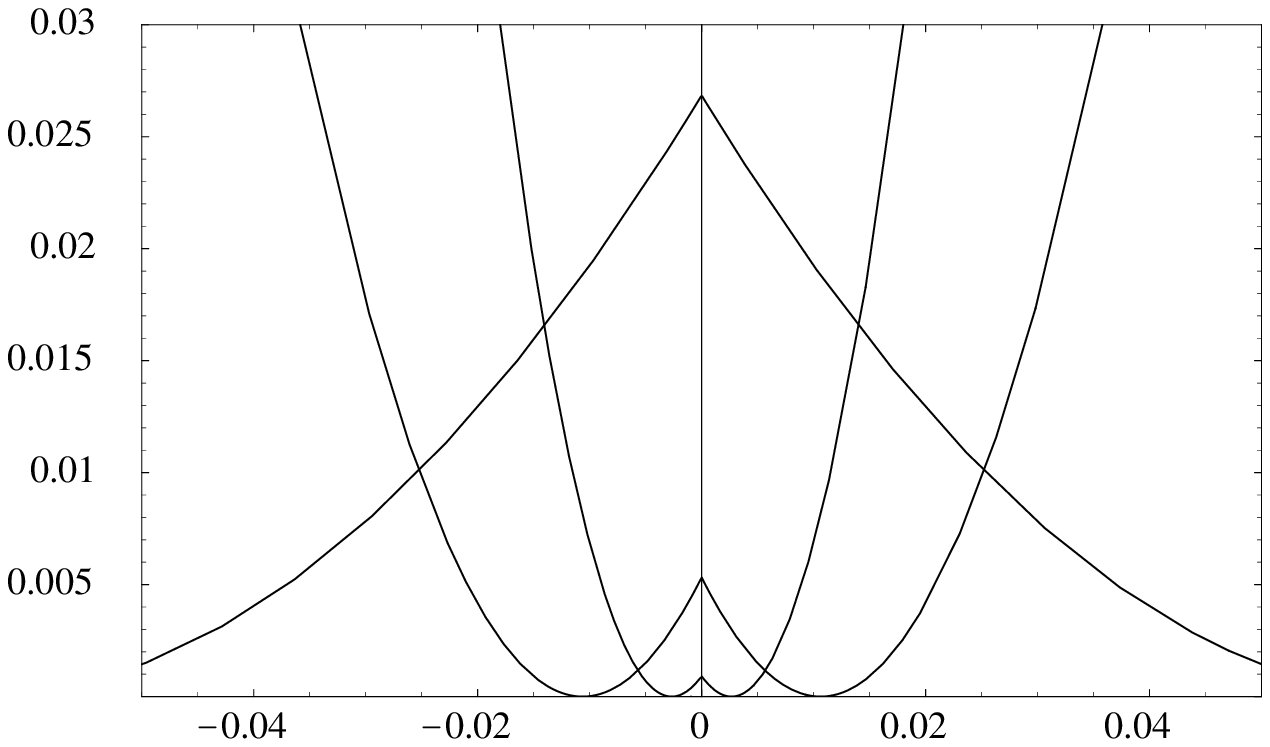,width=10cm}}
{\footnotesize\textbf{Figure 4:}  The same as Figure 3. 
The small portion of space near the
origin is shown in order to emphasize 
the suppression of the first three massive 
modes on the brane. The zero mode has value one, 
which is well out of the y axis 
range shown in the figure. }
\vspace{5mm}

Let us now compute the potential due to the exchange of all the modes 
between two static sources located on the brane.  
For this purpose we add a source 
term to (\ref{L5}), with a coupling to  the field $\Phi$ proportional to
$\sqrt{G_5} =M^{-3/2}$. 

Decomposing the five dimensional field in  four
dimensional modes, $\Phi=\sum_n\phi_n(y)\sigma_n(x^\mu)$,   
and integrating (\ref{L5})  over the extra compact dimension 
we  obtain the  following effective four 
dimensional Lagrangian:
\beqa \label{4dlag}
{\mathcal L}_4&=& \sum_{n,m=0}^\infty  \Bigg(\partial^\mu 
\sigma_m \partial_\mu
\sigma_n \left(\int {\mathrm d}y\,\phi_m\phi_n\right)
+  \left(\sigma_m \sigma_n\right) \left(\int {\mathrm d}y\, 
\partial^y\phi_m\partial_y\phi_n\right)  \nn \\
&&+  r_c  \left(\phi_m(0)\phi_n(0)\right) \partial^\mu
 \sigma_m \partial_\mu \sigma_n + \delta_{mn}\,\frac {\phi_n(0)} {M^{3/2}} 
\,\sigma_n\, \rho \Bigg).
\eeqa
Doing  the integrals with respect to the compact coordinate 
${\mathrm d} y$ one should take into account 
that the functions $\phi_n$ are \emph{not} orthogonal. 
Nevertheless, 
we can proceed as follows: 
The field equation for  $\phi_m$ can be written as: 
\beq
\phi_m\left( \partial_y^2 + m^2_n + r_c m^2_n \delta(y) \right) \phi_n =0.
\eeq
On the other hand, 
\beq
\phi_n\left( \partial_y^2 + m^2_m + r_c m^2_m \delta(y) \right) \phi_m =0.
\eeq
Integrating both equations w.r.t.  $y$ and subtracting them, we  obtain
\beq
0 = 
\left(m^2_n-m^2_m\right)\int d y \, 
\left(1+ r_c \delta(y)\right)\phi_n\phi_m.
\eeq
This implies that the integral on the right   
hand side vanishes unless $m=n$. 
Using this fact, we  obtain
\beq\label{trick}
\int d y\,\partial^y\phi_n\partial_y\phi_m = m_{n}^2
\int d y \, \phi_n\phi_m \left(1 + \delta(y) r_c\right) =m_{n}^2 
\delta_{mn} \left(1+ r_c |\phi_n(0)|^2\right).
\eeq
Furthermore, 
inserting    (\ref{trick}) in (\ref{4dlag}) we get
\beq 
{\mathcal L}_4= \sum_{n=0}^\infty
\Big( \left(1+r_c\left|\phi_n(0)\right|^2\right)\left(\left(\partial^\mu 
\sigma_n\right)^2 
+ m_n^2 \left(\sigma_n \right)^2 \right)
 + \frac {\phi_n(0)} {M^{3/2}} \sigma_n \rho\Big).
\eeq
In order to normalize canonically the  kinetic terms we   absorb the factor 
$(1+  r_c  |\phi_n(0)|^2)^{1/2} $  into the  redefinition of the  
fields. Thus, the low energy Lagrangian   becomes  
\beq \label{L4}
{\mathcal L}_4= \sum_{n=0}^\infty\Big( \left(\partial^\mu 
\sigma_n\right)^2 + 
m_n^2 \left(\sigma_n \right)^2  + \frac {1} {\sqrt{1 + 
r_c|\phi_n(0)|^2}
}\frac{\phi_n(0)} {\sqrt{ M^{3}}}  \sigma_n \rho\Big).
 \eeq
From (\ref{L4}) we  can read off the couplings of the various modes to 
four dimensional matter. In the case of   $r_c\gg R$, we can approximate
$m_n \simeq (2n-1)\pi/R$. Using this and some simplifying algebra 
we derive: 
\begin{eqnarray}
G_0 & = & \frac{1}{M^3} \frac{1}{r_c+ R} \simeq  \frac{1}{r_c 
M^3}
 \equiv G_\infty,  \label{zeromode1} \\
G_n & = & \frac{1}{r_c M^3} 
\left(\frac{R}{r_c}\right)\frac{2}{[(2n-1)
\pi/2]^2 + R/r_c + (R/r_c)^2}\nonumber \\
& \simeq& \frac{1}{r_c M^3}\left(\frac{2}{\pi^2}\right)\left(\frac{R}
{r_c}\frac{1}{n^2}\right). \label{massive1}
\end{eqnarray}

An equivalent  way to extract the information about  the potential 
is to compute the propagator of the field $\Phi$ 
from the \mbox{Lagrangian (\ref{L5})}. For this we have to solve the 
following equation (for simplicity we work in Euclidean space)
\beq
\left[\frac{1}{r_c} \partial^A\partial_A + \delta(y) \partial
^\mu\partial_\mu\right]\Delta(x, y)=
 -\delta^4(x)\delta(y),
\eeq
where $\Delta$ has  dimension two.
The procedure is similar to that used in \cite{DGPind}
with the only difference  being 
that the fifth dimension is now compact. 
We turn  to  4-dimensional momentum  space 
\beq
\left[\frac{1}{r_c}\left(\partial^2_y - p^2\right) -\delta(y) p^2 \right]
\Delta(p,y)
 = -\delta (y),
\eeq
and take the ansatz $\Delta(p,y)= D(p, y) B(p)$,  with $D(p, y)$ 
satisfying
\beq \label{Deq}
\left(\partial^2_y - p^2\right)D(p, y) = -\delta(y).
\eeq
This gives 
\beq\label{B}
B(p)= \frac{r_c}{1 + p^2 D(p, 0)~}.
\eeq
Eq.  (\ref{Deq}) can be solved in the compact space by 
expanding both sides in Fourier modes
\beq 
D(p,y)=\sum_{n=-\infty}^{+\infty}e^{i\frac{2\pi n}{R}y}D_n(p) \qquad 
\delta(y)=\frac{1}{R}\sum_{n=-\infty}^{+\infty}e^{i\frac{2\pi n}{R}y}. 
\eeq
As a result, one finds: 
\beq
D_n = \frac{1}{R} \frac{1}{p^2+(2\pi n /R)^2}.
\eeq 
To calculate the force between sources localized on 
the brane, we need to evaluate the propagator at $y=0$. Using
the expression  
\beq
D(p,0) = \sum_{n=-\infty}^{+\infty}D_n(p)= \frac{1}{p^2 R} \left[ 
\frac{pR}{2}\, \coth\left(\frac{pR}{2}\right)\right],
\eeq
we derive 
\begin{eqnarray} \label{propagator1}
\Delta (p, 0)&=& \frac{1}{p^2}\frac{r_c}{R}\frac{(pR/2)\coth(pR/2)}{1+
 (r_c/R)(pR/2)\coth(pR/2)} \nonumber \\  
&=& \frac{1}{p^2}\left[\frac{1}{1+ (1/pr_c)2\tanh(pR/2)}\right].
\end{eqnarray} 
The propagator never deviates substantially from $1/p^2$ over
the whole range of $p$. The large-$p$  and small-$ p$ behavior are 
respectively
\begin{eqnarray} \label{variation1}
\Delta(p,  0) &\simeq&  \frac{1}{p^2}\left[1-\frac{2}{pr_c}\right] 
\qquad p\,R\gg 1 \; \textrm{(short distances)} \nonumber \\
\Delta(p,  0) &\simeq &  \frac{1}{p^2}\left[1 - \frac{R}{r_c}\right]
 \qquad p\,R\ll 1 \; \textrm{(large distances)}.
\end{eqnarray} 
The maximal  deviation in  the coefficient of $1/p^2$ is of order $R/r_c$ 
as before. If we continue this expression to 
Minkowskian space $p\to ip_M$, the propagator becomes 
\beq
\Delta(p, 0)  = - \frac{1}{p_M^2}\left[\frac{1}{1+ 
(1/p_Mr_c)2\tan(p_MR/2)}\right]~. 
\eeq
The poles in this propagator are located 
at $r_c p_M= -2 \tan(p_MR/2)$, in agreement with
equation (\ref{mass})~.

\end{document}